\documentstyle[twocolumn,aps,epsfig,floats]{revtex}

\begin{document}

\draft
\tolerance = 10000

\setcounter{topnumber}{1}
\renewcommand{\topfraction}{0.9}
\renewcommand{\textfraction}{0.1}
\renewcommand{\floatpagefraction}{0.9}

%Fixing abstract in twocolumn mode
\twocolumn[\hsize\textwidth\columnwidth\hsize\csname
@twocolumnfalse\endcsname

\title{Globally coupled bistable elements as a
model of group decision making}
\author{Dami\'an H. Zanette}
\address{\it Consejo  Nacional de Investigaciones  Cient\'{\i}ficas y
T\'ecnicas\\ Centro At\'omico Bariloche and Instituto Balseiro\\ 8400
Bariloche, R\'{\i}o Negro, Argentina}
\maketitle

\begin{abstract}
A simple  mathematical model  is proposed  to study  the effect of the
average trend of a population on the opinion of each individual,  when
a group decision has to be made  by voting.  It is shown that  if such
effect is strong enough a  transition to coherent behavior occurs,  in
which  the  whole  population  converges  to  a  single  opinion. This
transition has the character of a first-order critical phenomenon.
\end{abstract}

\vspace{1cm}

%Fixing abstract in twocolumn mode
]

Group  decision  making  is  a  complex  social  process  in which the
inherent factors that determine the position of each individual --such
as  previous  experience,   prospective  benefits,  current   personal
circumstances, and character-- interact  in a nontrivial way  with the
average trend, to which individuals are exposed through  communication
between them. Group decision results from this interaction as an
emerging property of their collective behavior.

Consider as a  specific case an  ensemble of individuals  that have to
choose, at a given  time in the future  and by individual voting,  and
option among a prescribed set of instances.  After vote counting,  the
decision is  simply taken  for the  most voted  option.  This decision
--which, once votes have been emitted, is straightforwardly  defined--
results however from the complex collective process that builds up the
opinion  of  each  individual  \cite{opin}.   During  a certain period
previous to the voting act, in fact, the opinion of a given individual
evolves due to the modulation that the knowledge of another's position
imposes  on  the  own  tendency.   In  an  efficiently   communicating
ensemble, like in any modern population, individuals are  continuously
exposed to the average opinion of the ensemble --for instance, through
poll results, published by mass communication media-- and are expected
to be  more or  less strongly  influenced by  this collective  element
\cite{mass}.

This  paper  is  aimed  at  exploring,  in  the  frame  of  a   simple
mathematical model, how the effects  of the personal trend and  of the
average  opinion  in  defining  the  individual vote combine with each
other to lead the  group to its collective decision.  For the  sake of
concreteness,  suppose  that  the  population  has to choose by voting
among  two  candidates,  $C^+$  and  $C^-$.   In  the  model, the time
evolution of the  opinion of the  $i$-th individual is  described by a
variable $x_i(t)$, with $x_i \in [-1,1]$  for all $i$ and $t$.   Large
values of $x_i$, $|x_i | \approx 1$, are to be associated with a  firm
decision to vote for one of the two candidates, ($C^+$ for $x_i>0$ and
$C^-$ for $x_i<0$, say), whereas small values of $x_i$ correspond to a
looser opinion.   In any  case, when  the voting  act takes  place the
individual opinion is quenched and  the decision is made according  to
the sign of $x_i$ at that  time. In practice, it is supposed  that the
typical evolution times  for the individual  opinion are shorter  than
the time  elapsed up  to the  voting act,  so that  one will focus the
attention on the long-time asymptotics of the model.

The average opinion, which is expected to play a relevant role in  the
definition of the  individual decision, is  here characterized by  the
arithmetic mean value
\begin{equation}    \label{mean}
\bar x (t) =\frac{1}{N} \sum_i x_i(t),
\end{equation}
where $N$ is the size of the population. This mean value is a  measure
of how much defined is the global trend towards on the two candidates.
In fact, the sign of $\bar x$ at the time of the voting act determines
the chosen candidate.

To stress the effect of the average opinion on the individual vote  it
is assumed that, in the absence of such effect, each individual  would
simply reinforce his or  her initial personal opinion as time elapses.
This means, in  particular, that a  given individual would  not change
his  or  her  original  preference  for  one  of  the candidates. This
behavior is well represented  by the following dynamical  equation for
$x_i(t)$:
\begin{equation}    \label{indiv}
\frac{dx_i}{dt} =x_i-x_i^3 \ \ \ \ \ (i=1,2,\dots ,N).
\end{equation}
In fact, the solution to this equation approaches the asymptotic value
$x_i(\infty)=+1$ or  $x_i(\infty)=-1$ depending  on the  initial value
$x_i(0)$ being positive or negative, respectively. Moreover,  $x_i(t)$
does not change  its sign during  the whole evolution.  If $x_i(0)=0$,
then $x_i(t)=0$ for  all $t$, but  this stationary state  is unstable.
These facts can readily be verified from the explicit solution to  Eq.
(\ref{indiv}), which reads,
\begin{equation}    \label{sol}
x_i(t)= \frac{x_i(0)}{\sqrt{x_i(0)^2-[x_i(0)^2-1]\exp(-2t)}}.
\end{equation}
From  a  dynamical  viewpoint,  Eq.   (\ref{indiv})  implies that each
individual behaves as a  bistable element, its asymptotic  state being
fixed by  the initial  condition. In  physics, this  kind of model has
been  used  to  study  spin  systems  (in the soft-spin approximation)
\cite{spin} and neural networks \cite{nn}.

The effect of the average  opinion on the evolution of  the individual
trend is  described by  modifying Eq.  (\ref{indiv}) in  the following
way:
\begin{equation}    \label{model}
\frac{dx_i}{dt}=x_i-x_i^3 + k_i (\bar x - x_i)
\ \ \ \ \ (i=1,2,\dots ,N),
\end{equation}
where $\bar x  (t)$ has been  defined in (\ref{mean})  and $k_i$ is  a
constant that, as discussed  in the following, measures  the influence
of the average opinion on  the $i$-th individual. For $k_i>0$  the new
terms drive  $x_i(t)$ towards  the average  $\bar x(t)$.  In fact, for
large values  of $k_i$  and slowly  varying $\bar  x$, the  individual
variable $x_i$ would exponentially approach the average. The new terms
thus represent, for positive $k_i$,  a trend of the $i$-th  individual
to follow the average opinion,  which can either reinforce or  compete
with his or her individual  position.  Negative values of  $k_i$ would
correspond to individuals who tend to take a position opposite to  the
average.

In  a  physical  context,  the  new terms represent an ``interaction''
between individuals.   From a  mathematical viepoint,  in fact,  those
terms couple the  set of equations  (\ref{model}) through the  average
$\bar x $, which depends on  the whole set of $x_i$ $(i=1,\dots  ,N)$.
This coupling  makes it impossible to give the exact  solution to  the
model  equations  (\ref{model}),  and  the  system  has  to be treated
numerically. In particular, note that it is not possible to derive an
autonomous equation for the evolution of the average $\bar x(t)$.

The case where the coupling constant  $k_i$ is positive and the   same
for all  individuals, $k_i  = k>0$  for all  $i$, is considered first.
Obviously, for $k=0$ the uncoupled ensemble --whose behavior has  been
discussed above-- is recovered.  For $k=1$, Eq. (\ref{model})  reduces
to
\begin{equation}    \label{k1}
\frac{dx_i}{dt}=\bar x-x_i^3.
\end{equation}
Let  $r_{ij}=x_i-x_j$  be the difference between the states of any two
individuals.   It  can  be  shown   from  Eq.   (\ref{k1})  that,   if
$-1<x_i,x_j<1$, $r_{ij}$ tends to  zero as  time elapses  \cite{Z}. In
other words,  for $k=1$  the model  predicts that  all the individuals
will have the same opinion at sufficiently long times.
        
\begin{figure}
\begin{center}
\psfig{figure=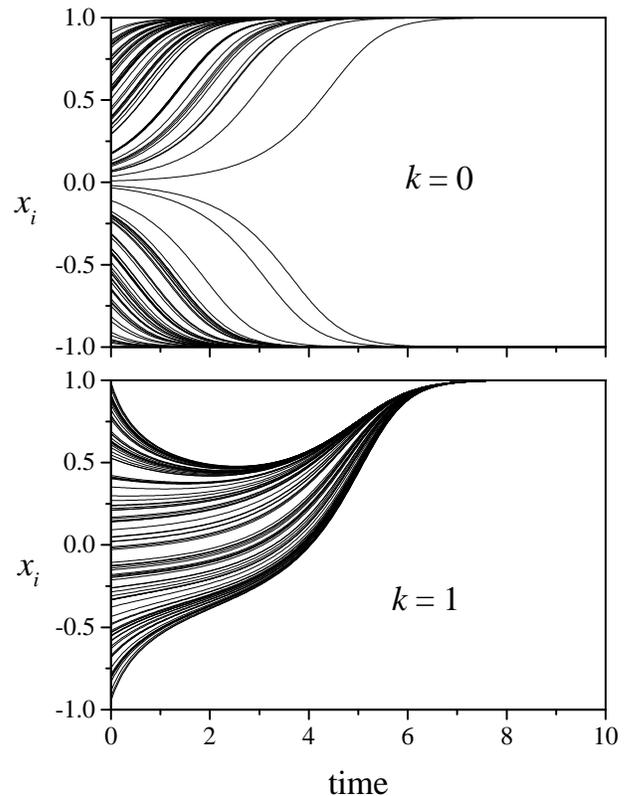,width=\columnwidth}
\end{center}
\caption{Evolution  of  $x_i$  for  a  population  of   1000
individuals with  $k=0$ and  $k=1$. For  clarity, only  100 states are
displayed. The initial distribution is uniform over $[-1,1]$.
}
\end{figure}

Figure  1  shows  the  evolution  of  $x_i$  in  a  population of 1000
individuals for  $k=0$ and  $k=1$. For  the sake  of clarity, only 100
variables are displayed. Initially, the individual states are randomly
distributed in $[-1,1]$.  As  expected, for $k=0$ the states  are soon
divided  into  two  clusters,  according  to  their  signs. For $k=1$,
instead, all states are attracted to a single cluster. Since for  this
value of the coupling constant the average opinion is already dominant
and all the individuals behave in a coherent way, it can be  predicted
that for $k>1$ the dynamics of the ensemble is qualitatively the same.
This is indeed verified from numerical results. On the other hand, for
$0<k<1$ a transition is expected to occur between the two  qualitative
different behaviors  observed at $k=0$ and  $k=1$. This  transition is
characterized in the following.

According to numerical calculations, for sufficiently small values  of
the coupling constant the collective behavior qualitatively reproduces
the  evolution  of  the  uncoupled  ensemble  ($k=0$). In fact, if the
initial distribution of $x_i$ is uniform over $[-1,1]$ the  population
becomes divided into two groups  --as when, for $k=0$, both  signs are
initially present. If, instead, one of the signs is much more abundant
than the other,  all the variables  may ultimately converge  to one of
the extreme values $x_i=\pm 1$ --as  when, for $k=0$ only one sign  is
initially present. The interacting ensemble is therefore ``bistable,''  
in
the  sense  that  two  qualitatively  different  asymptotic states are
observed depending on the  initial condition:  either  all individuals
behave coherently,  or they  become divided  into two  groups. On  the
other hand, as  stated above, for  larger values of  $k$ only coherent
behavior  is  observed.  Figure  2  illustrates  these  behaviors  for
intermediate values of $k$.
         
\begin{figure}
\begin{center}
\psfig{figure=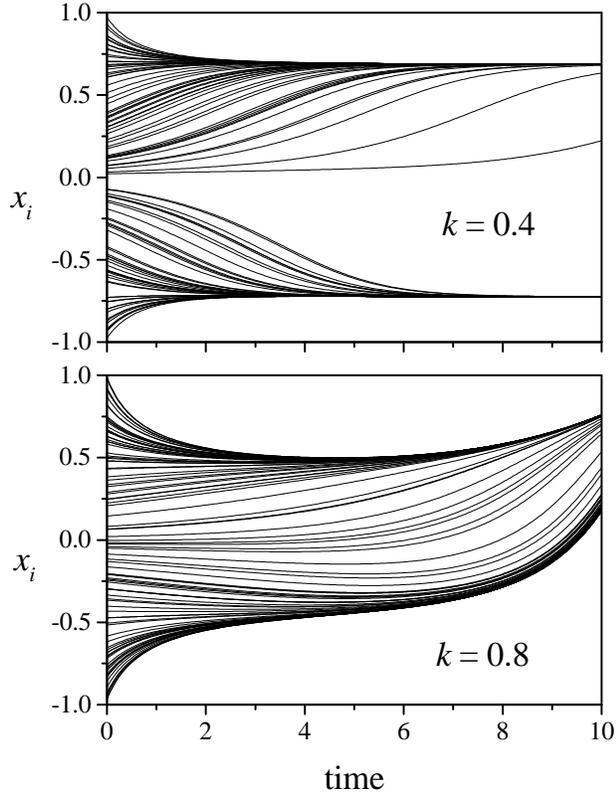,width=\columnwidth}
\end{center}
\caption{Evolution  of  $x_i$  for  a  population  of   1000
individuals with $k=0.4$ and $k=0.8$. For clarity, only 100 states are
displayed. The initial distribution is uniform over $[-1,1]$.
}
\end{figure}

The transition  between bistable  and coherent  behavior is  due to  a
stability  change  in  the  possible  asymptotic states of the coupled
ensemble. Suppose  that, as  the system  evolves, the  $N$ individuals
become divided  into two  groups. One  of them,  with $pN$ individuals
($0\le p\le  1$) approaches  the asymptotic  state $X_1$,  whereas the
other,  with  $(1-p)N$  individuals,  approaches  $X_2$.  It has to be
stressed that  the value  of $p$  depends in  a nontrivial  way on the
initial  condition,  and  cannot  be  analytically  determined  {\it a
priori}.  According  to  Eq.  (\ref{model})  the  following identities
should hold for $t\to \infty$ and $N\to \infty$:
\begin{equation}    \label{eqs}
\begin{array}{rl}
0&= (1-k)X_1+k[pX_1+(1-p)X_2]-X_1^3 \\   \\
0&= (1-k)X_2+k[pX_1+(1-p)X_2]-X_2^3.
\end{array}
\end{equation}
These equations  include also  the case  of coherent  behavior, if one
puts $X_1=X_2$ with any value  of $p$. Their solutions constitute  the
set  of  stationary  states  for  the  system,  whose stability can be
studied by means of standard linearization around equilibria.

Equations  (\ref{eqs})  can  be  reduced  to  a  9th-degree polynomial
equation for either $X_1$ or $X_2$, and have therefore nine  solutions
--which  in  general  are  complex  numbers.   The  trivial   solution
$X_1=X_2=0$ is unstable. The remaining eight solutions can be  grouped
into symmetrical pairs, ($X_1,X_2$)  and ($-X_1,-X_2$), both with  the
same stability  properties.   It is  therefore enough  to analyze, for
instance,  the  four  solutions  with  $X_1\ge  0$. (i) The first one,
$X_1=X_2=1$, is stable for all  $k$ and corresponds to the  asymptotic
state of  coherent evolution.  (ii) The  second solution  is real  and
unstable for all $k$. It approaches the unstable solution $(0,-1)$ for
$k  \to  0$  and  the  trivial  solution  for  $k\to 1$. (iii) Another
unstable solution approaches the  unstable state ($0,1$) as  $k\to 0$.
(iv) Finally, there is a  stable solution that approaches $(1,-1)$  as
$k\to 0$. This solution corresponds to the case where the  individuals
have become divided into two groups.
                                              
\begin{figure}
\begin{center}
\psfig{figure=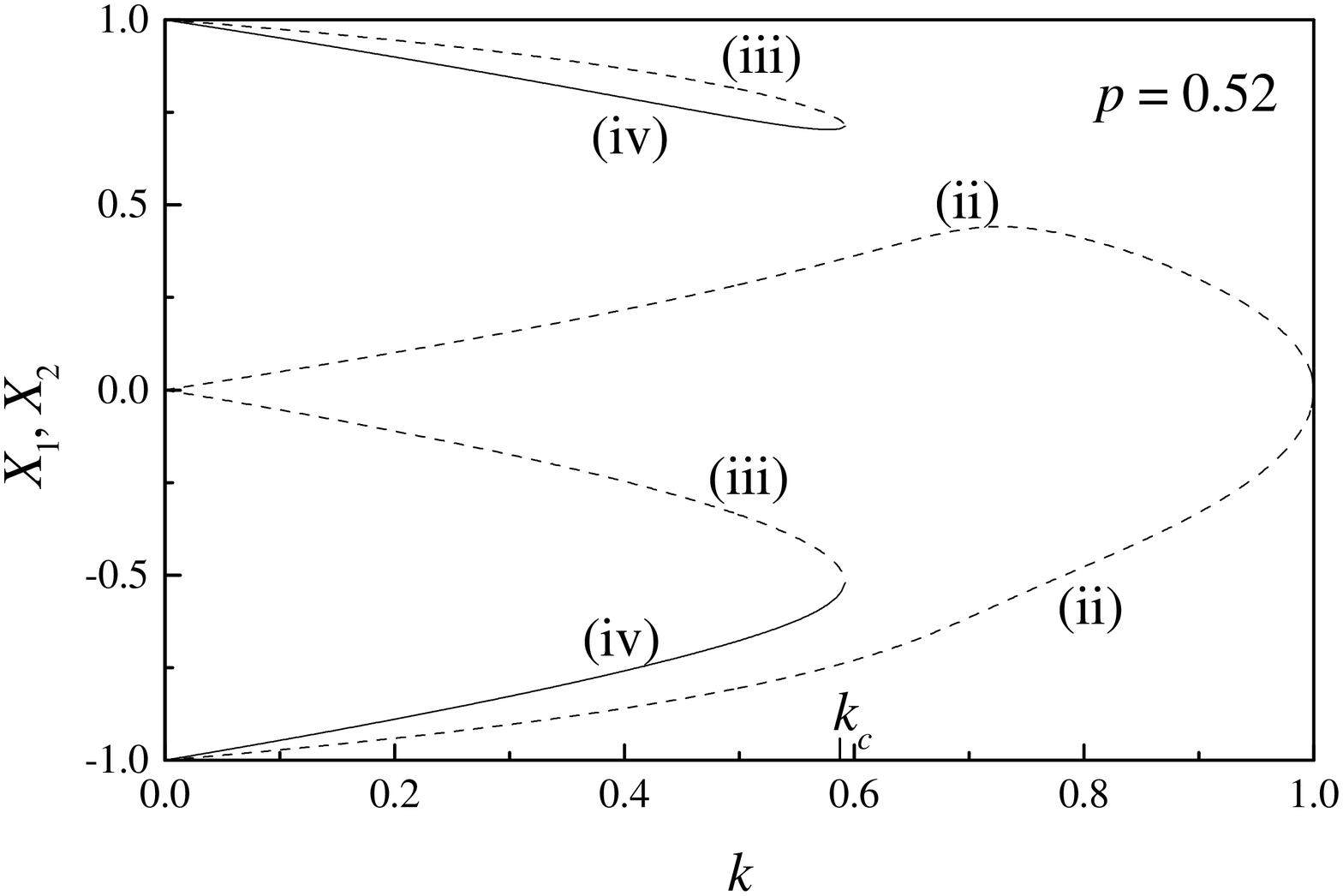,width=\columnwidth}
\end{center}
\caption{The equilibrium  states $X_1$ and $X_2$ as  a function
of $k$ for  solutions (ii), (iii),  and (iv) (see  text), at $p=0.52$.
For the  solutions displayed,  $X_1$ is  always positive  and $X_2$ is
always negative.
}
\end{figure}

Figure 3 shows the numerical results for $X_1$ and $X_2$ as a function
of $k$, for  $p=0.52$. Solid lines  indicate stable solutions  whereas
dashed lines stand  for unstable solutions.  As the coupling  constant
grows, there is a critical value $k_c$ at which the two solutions (iii)
and  (iv)  collide  and  become  complex.  At this critical value, the
solution where the population  is divided into two  groups dissapears.
The  value  of  $k_c$  is  related  to $p$ by the following polynomial
equation:
\begin{equation}    \label{kcp}
4k_c-18k_c^2 (p-p^2)+27 k_c^4(p^2-2p^3+p^4)=1 .
\end{equation}
Thus, for a given  value of $p$ --which  is determined by the  initial
condition--  and  $k<k_c$, two  qualitatively  different behaviors can
occur. Either $X_1=X_2=\pm 1$,  and the system evolves  coherently, or
$X_2\neq X_1$, and  the individuals are  divided into two  groups. For
$k>k_c$,  instead,  only  coherent  behavior  is  possible.  Thus, for
sufficiently large $k$, the opinion of the whole population approaches
the same state. Figure 4 shows  a phase diagram $k$ versus $p$,  where
the boundary between the zones  of bistability and coherence given  by
Eq. (\ref{kcp}) is shown.

\begin{figure}
\begin{center}
\psfig{figure=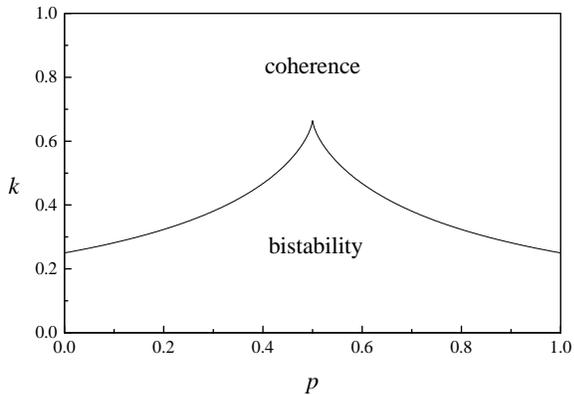,width=\columnwidth}
\end{center}
\caption{Phase diagram in the ($p,k$) plane, displaying the
zones of coherent behavior and bistability.
}
\end{figure}

The transition between bistability and coherence can be  characterized
by  a  single  order  parameter  introducing,  for  instance, the mean
difference $\delta$ between the states of any pair of individuals,
\begin{equation}    \label{delta}
\delta= \frac{1}{2} p(1-p)|X_1-X_2|,
\end{equation}
which has been plotted  in Fig. 5 as  a function of $k$  for $p=0.52$.
The  dependence  of  $\delta$   on  the  coupling  constant   suggests
classifying the transition as a first-order critical phenomenon.

Naturally, the assumption that  all the individuals in  the population
are equally  influenced by  the average  opinion --i.e.  that all  the
individuals  have  the  same  coupling  constant  $k$--  is not a very
realistic  one.  Rather,  it  is  to  be  expected  that  the coupling
constants  are  distributed  within  a  certain  interval,  with  some
individuals being more influenced by another's opinion than other.  It
could moreover be supposed that some individuals are {\it  negatively}
affected  by  the  mean  trend,  tending  to  make  their  own opinion
diverging from the average. This case would correspond to $k_i<0$.

In this  case of  inhomogeneous behavior  it can  be easily shown from
Eq.  (\ref{model})  that the asymptotic  state of $x_i(t)$  depends on
the value of $k_i$. This relation is implicitly given by the equation
\begin{equation}    \label{ki}
k_i=\frac{x_i^3-x_i}{\bar x-x_i}.
\end{equation}
Numerical simulations show however that, as far as the distribution of
coupling  constants  $k_i$  is  moderately  narrow,  the   qualitative
collective behavior is the same as for uniform $k$. Mathematically, it
is  not  an  easy  task  to  characterize  the situation in which this
behavior breaks  down  as  the values  of $k_i$  become more  and more
scattered. Is is nevertheless expected that coherent evolution can  be
destroyed if  the coupling  constants are  sufficiently different from
each other, including in particular some negative values. A sufficient
condition for coherence  to fail is  in fact that  a single individual
has a  coupling constant $k_i<-2$. In  this situation,  however, it is
this  only  individual  who  fails  to  behave  coherently,  thus  not
affecting the result of the collective decision making.

\begin{figure}
\begin{center}
\psfig{figure=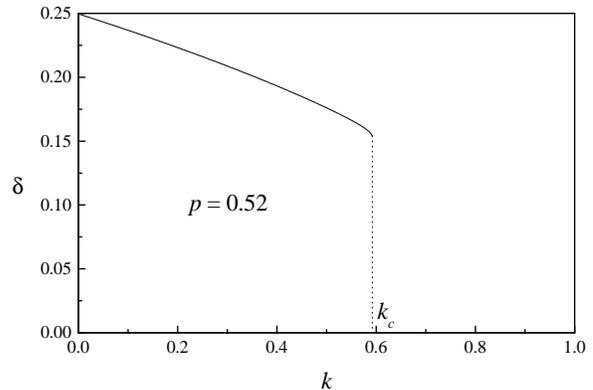,width=\columnwidth}
\end{center}
\caption{The order parameter $\delta$ as a function of $k$, for
$p=0.52$. At $k_c \approx 0.592$, $\delta$ vanishes suddenly.
}
\end{figure}
In summary, it has been here shown within a simple mathematical  model
that, under a sufficiently strong influence of the average trend of  a
population on the opinion  of each individual, the  population behaves
coherently and votes converge  towards a single candidate.  For weaker
coupling between individuals, instead,  votes are more evenly  divided
between the two candidates.  The transition between both  behaviors is
abrupt and, in fact, has the character of a critical phenomenon.  This
is  qualitatively  similar  to  the  ferromagnetic  phase   transition
observed  in  spin  systems  --though  in this physical phenomenon the
phase transition is of  the second order \cite{ferro}.  Indeed, beyond
the  critical  point  the  state  of  all  the  elements in the system
coincide, even in  spite of the  initial condition corresponding  to a
uniform distribution of states. The coupling mechanism is in fact able
to break  the initial  macroscopic homogeneity,  enhancing microscopic
fluctuations. It  can be  interesting to  further analyze  this model,
including for instance local communication ways between individuals as
well as noise, that can perturb in a nontrivial way the properties  of
the quoted transition \cite{VK}.

The present  results could  encourage unfair,  unscrupulous candidates
to manipulate poll results published in mass media --if they have  the
power to do so-- in their own benefit.


\begin{references}
%\vspace{-1.2cm}
%
\bibitem{opin} B. Berenson, P. Lazarsfeld, and W. McPhee, {\it Voting.
A  Study  of  Opinion  Formation  in  a  Presidential  Campaign}  (The
University of Chicago Press, Chicago, 1954).

\bibitem{mass} J.  Klapper, {\it  The Effects  of Mass  Communication}
(The Free  Press, New  York, 1960);  E. Katz,  {\it The Utilization of
Mass Communication  by the  Individual} (Oxford  University Press, New
York, 1979).

\bibitem{spin} P. Jung, U. Behn, E. Pantazelou, and F. Moss, Phys.
Rev. A {\bf 46}, R1709 (1992).

\bibitem{nn} H. Sompolinsky, Phys. Rev. A {\bf 34}, 2571 (1986); {\it
ibid.} {\bf 37}, 4865 (1988).

\bibitem{Z} D.H. Zanette, Phys. Rev. E {\bf 55}, 5315 (1997).

\bibitem{ferro} D.L. Goodstein, {\it States of Matter} (Dover, New
York, 1975).

\bibitem{VK} N. van Kampen, {\it Stochastic Processes in Physics and
Chemistry} (North-Holland, Amsterdam, 1992).
\end{references}
\end{document}